\documentstyle[12pt,epsfig]{article}
\def\bmq{{\bf q}}
\def\bmr{{\bf r}}

\def\minus{\mbox{$-$}}
\begin{document}
\vspace*{-1.0in}
\hspace{\fill} \fbox{\bf LA-UR-98-5728}

\begin{center}

{\Large {\bf Hadronic Vacuum Polarization and the Lamb Shift}}\\

\vspace*{0.80in}

J.L.\ Friar \\
Theoretical Division \\
Los Alamos National Laboratory \\
Los Alamos, NM  87545 USA\\

\vspace*{0.15in}

and

\vspace*{0.15in}

J.\ Martorell\\
Departament d'Estructura i Constituents de la Materia\\
Facultat F\'isica\\
Universitat de Barcelona\\
Barcelona 08028 Spain\\

\vspace*{0.15in}

and

\vspace*{0.15in}

D.\ W.\ L.\ Sprung\\
Department of Physics and Astronomy\\
McMaster University\\
Hamilton, Ontario, L8S 4M1 Canada\\

\end{center}

\vspace*{0.50in}

\begin{abstract}
Recent improvements in the determination of the running of the fine-structure
constant also allow an update of the hadronic vacuum-polarization contribution
to the Lamb shift.  We find a shift of \minus3.40(7) kHz to the 1S level of
hydrogen. We also comment on the contribution of this effect to the
determination by elastic electron scattering of the r.m.s. radii of nuclei.
\end{abstract}
\pagebreak

Hadronic vacuum polarization (VP)\cite{1,2,3,3x} contributes to several
quantities that have elicited much recent interest:  g-2 of the muon, the
effective fine-structure constant at the energy scale of the Z-boson mass, the
hyperfine splitting in muonium and positronium\cite{4}, and the energy levels of
the hydrogen atom\cite{5}.

\begin{figure}[htb]
\epsfig{file=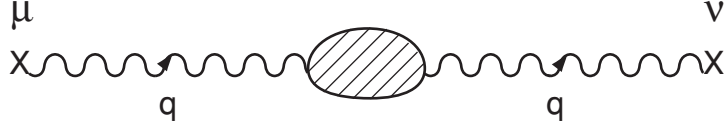,bbllx=97pt,bblly=544pt,bburx=367pt,bbury=597pt}

\caption{Vacuum polarization insertion into a virtual photon propagator.}
\end{figure}

The shaded ellipse in Figure (1) represents the creation and subsequent
annihilation of arbitrary hadronic states by virtual photons on the left (with
polarization, $\mu$) and right (with polarization, $\nu$).  For (squared)
momentum transfers, $q^2$, comparable to the masses of various components in the
shaded ellipse, the screening of charge that defines the vacuum polarization
changes with $q^2$ and leads to an effective fine-structure constant that
depends on $q$: $\alpha(q^2)$.  If the entire unit in Fig.\ (1) is inserted as a
vertex correction on a lepton, it will also affect g-2 of that lepton via an
integral over $q^2$.  Finally, at very small values of $q^2$, inserting the unit
between an electron and a nucleus will lead to a shift in hydrogenic energy
levels.

The S-matrix for the process sketched in Fig.\ (1) (with one VP insertion) has 
the form (in a metric where $p^2 = m^2$)

\begin{equation}
S^{\mu\nu} = \left(\frac{-i e_0 \, g^{\mu\alpha}}{q^2}\right) 
[i\pi^{\alpha\beta}(q^2)] 
\left(\frac{-i e_0 \, g^{\beta\nu}}{q^2}\right) \ \ ,  
\end{equation}

\noindent
and for the process without polarization (just a single photon propagator) is 
$(-i e_0^2 \, g^{\mu\nu}/q^2)$. The strength is determined by the bare electric
charge, $e_0$, located at the two ends of Fig.\ (1).  The polarization structure
function $\pi^{\alpha\beta}(q^2)$ must be gauge invariant, which requires

\begin{equation}
\pi^{\alpha\beta}(q^2) = -(g^{\alpha\beta}q^2 - q^\alpha q^\beta) \pi(q^2) \ \ .
\end{equation}

\noindent
Coupling the $q^\alpha q^\beta$ term to any conserved current (e.g., an electron
or nucleus) leads to vanishing results and simplifies the tensor structure of
$S^{\mu\nu}(\sim g^{\mu\nu})$.  The sequence of 0, 1, 2, $\cdots$ insertions 
of $\pi^{\alpha\beta}$ into a photon propagator (Fig.\ (1) shows one insertion) 
generates the geometric series
\begin{equation} 
S^{\mu\nu} = \frac{-i g^{\mu\nu} e^2_0}{q^2}(1 - \pi + \pi^2 \cdots) 
= \frac{-i g^{\mu\nu}}{q^2}\frac{e^2_0}{1 + \pi(q^2)} \ \ .
\end{equation}

\noindent 
Clearly, at $q^2 = 0$ we expect $e^2_0/(1 + \pi(0))$ to be $e^2$, the
renormalized charge (squared).  Since $\pi$ itself is proportional to $e^2_0$, 
we form $\Delta\pi(q^2) = \pi(q^2) - \pi(0)$, rearrange, and find to first order
in $e^2$

\begin{equation}
S^{\mu\nu} = \frac{-i g^{\mu\nu}}{q^2} e^2 (q^2) \ \ ,
\end{equation}

\noindent
where

\begin{equation}
e^2(q^2) = \frac{e^2}{1 + \Delta\pi(q^2)} 
\end{equation}

\noindent
is the (squared) effective charge at the scale $q^2$.  Much recent work has been
devoted to the numerical determination\cite{1,2,3,3x} of $\frac{e^2(q^2)}{4\pi} 
\equiv \alpha(q^2)$.

Our purpose is to treat vacuum polarization in hydrogenic ions with nuclear 
charge $Z e$. The momentum-space Coulomb potential generated by Eq.\ (5) is 
given by

\begin{equation}
V(q^2) = \frac{Z e^2(q^2)}{q^2} = \frac{-Z e^2(q^2)}{\bmq^2} \cong 
\frac{-Z e^2}{\bmq^2} - Z e^2 \pi^{\prime} (0) \ \ ,
\end{equation}

\noindent
where $q^2 = - \bmq^2$ for this term in Coulomb gauge, and $\Delta \pi(q^2)
\cong q^2 \pi^\prime(0) + \cdots$, where the first term in Eq.\ (6) is the usual
Coulomb potential and the second is the vacuum polarization potential, which we 
write in configuration space as

\begin{equation}
V_{\rm VP}(\bmr) = -4 \pi (Z \alpha) \pi^{\prime} (0) \delta^3(\bmr) \ \ .
\end{equation}

\noindent
Thus for the n$^{th}$ S-state the energy shift is given by

\begin{equation}
E_{\rm VP} \cong -4 \pi (Z \alpha) \pi^\prime(0)|\phi_n(0)|^2 = 
\frac{-4(Z \alpha)^4\pi^\prime(0)\mu^3}{n^3} \ \ ,
\end{equation}

\noindent
where $\mu$ is the hydrogenic reduced mass.  The intrinsic nature of vacuum
polarization is attraction, so we expect $\pi^\prime(0) > 0$.

By slicing Fig.\ (1) through the polarization insertion, we can reexpress
$\pi(q^2)$ as a dispersion relation, with an imaginary part proportional to
$\sigma_h(q^2)$, the cross-section for producing all hadron states in
$e^+e^-$ collisions.  A single subtraction then produces $\Delta \pi (q^2)$ in
the form

\begin{equation}
\Delta \pi(q^2) = \frac{q^2}{\pi} \int_{4m^2_\pi}^{\infty} \frac{dt\, Im(\Delta 
\pi  (t))}{t (t - q^2 - i\epsilon)} \ \ ,
\end{equation}

\noindent
where $Im(\Delta \pi(t)) = t\, \sigma _h (t)/(4 \pi \alpha (t) )^2$.  Utilizing 
all available $e^+e^-$ collision data (and some theory) permits an accurate
interpolation of $\sigma_h(t)$, and $\Delta \pi(q^2)$ can be constructed
numerically\cite{1,2,3,3x}.

We require only $\pi^\prime(0)$ for the hadronic VP (henceforth subscripted with
$h$), which is given by the parameter $\ell_1$ in section 1.5 and the error from
Figure (7) of Ref. \cite{1} (note that our $\Delta \pi = - \Delta \alpha$ of
\cite{1}):

\begin{equation}
\pi_h^\prime (0) = 9.3055 (\pm 2.2\%) \cdot 10^{-3}\, {\rm GeV}^{-2} \ \ .
\end{equation}

\noindent
Equation (8) also applies to muon-pair vacuum polarization, for which 
$\pi^\prime_\mu (0) = \alpha/(15\pi m^2_\mu)$, where $m_\mu$ is the muon mass. 
We therefore obtain

\begin{equation}
\pi^\prime_h (0) = 0.671(15)\, \pi^\prime_\mu (0) \equiv \delta_h \,
\pi^\prime_\mu (0) \ \ ,
\end{equation}

\noindent
and thus

\begin{equation}
E^{\rm had}_{\rm VP} = 0.671(15)\, E^\mu_{\rm VP} \ \ ,
\end{equation}

\noindent
where for S-states Eq.\ (8) gives

\begin{equation}
E^\mu_{\rm VP} = \frac{-4\alpha (Z \alpha)^4\mu^3}{15 \pi n^3 m^2_\mu} \ \ ,
\end{equation}

\noindent
and a numerical value of \minus5.07 kHz for the 1S state of hydrogen. The much 
heavier $\tau$-lepton analogously contributes $\minus$0.02 kHz.

Previous values obtained for $\delta_h$ are displayed together with our value 
in Table I. Additional values were calculated in Refs. \cite{9} and \cite{5}.
The latter estimate used only the $\rho$-meson contribution\cite{4}, which is 
known to give the largest fractional contribution to $\pi_h^\prime(0)$, and
results for g-2 of the muon are consistent with that fraction ($\sim$ 60\%)
\cite{1}.  All tabulated values are consistent with the more accurate Eq.\ (12).
We also note that a noninteracting pion pair generates only $\sim$ 10\% of the
total hadronic contribution.

\begin{table}[htb]
\centering
\caption{Values of $\delta_h$.}
\hspace{0.25in}

\begin{tabular}{|c|cccc|}
\hline
{}&Ref. \cite{6} & Ref. \cite{7} & Ref. \cite{8} & This work \\ \hline
$\delta_h $  \rule{0in}{2.5ex}& 0.68 & 0.719(54) & 0.659(26) & 0.671(15) \\ 
\hline
\end{tabular}
\end{table}

Finally, we repeat a caveat from Ref. \cite{10}.  In elastic-electron-scattering
determinations of nuclear form factors (and hence their radii), the
radiative-corrections procedure \cite{11,12} that is used to analyze the data
corrects for $e^+e^-$ vacuum polarization, sometimes for the muon one, but 
typically not for the hadronic one. If one type of vacuum polarization is
omitted, equation (5) then demonstrates that the {\bf effective} measured form
factor expressed in terms of $F_0$ (the true form factor) is

\begin{equation}
F_{\rm eff}(q^2) = \frac{F_0(q^2)}{1 + \Delta \pi (q^2)} \ \ ,
\end{equation}

\noindent
and hence the effective radius is

\begin{equation}
\langle r^2 \rangle ^{1/2}_{\rm eff} = [\langle r^2 \rangle_0 - 
6\pi^\prime (0)]^{1/2} \ \ ,
\end{equation}

\noindent
where $6\pi_h^\prime(0) = 0.0022\, {\rm fm}^2$, for example.  Although this is a
very tiny effect, comparing a measured radius ($\langle r^2 \rangle_{\rm
eff}^{1/2}$ in Eq.\ (15)) with one determined from optical measurements {\bf
corrected for hadronic VP} (i.e., $\langle r^2 \rangle ^{1/2}_0$ from $F_0
(q^2)$) would be inconsistent.

In summary, we have updated the calculation of the hadronic vacuum polarization
correction in hydrogen using a recently obtained and more accurate value\cite{1}
for $\pi_h^\prime(0)$.  This leads to a shift of \minus3.40(7) kHz in the 1S 
level of hydrogen.  We also noted that elastic electron scattering from nuclei 
is not corrected for hadronic VP.

\pagebreak

\noindent {\bf Acknowledgements}

The work of J.\ L.\ F.\ was performed under the auspices of the United  
States Department of Energy. D.\ W.\ L.\ S.\ is grateful to NSERC Canada
for continued support under Research Grant No.\ SAPIN-3198. The work of 
J.\ M.\ is supported under Grant No.\ PB97-0915 of DGES, Spain. One of us
(J.\ L.\ F.) would like to thank P.\ Mohr of NIST for a stimulating series of 
conversations, and L.\ Maximon of The George Washington Univ. for information 
about radiative corrections.

\end{document}